\documentclass{ws-procs9x6}

\begin{document}

\title{The CRESST dark matter search}

\author{B. Majorovits , C. Cozzini, S. Henry, H. Kraus,  
V.~Mikhailik, A.J.B.~Tolhurst, D.~Wahl}

\address{Dept. of Physics, University of Oxford, Keble Road, OX1 3RH, England, 
\\email: majorovits@physics.ox.ac.uk}

\author{Y. Ramachers}

\address{University of Warwick, England}  

\author{G. Angloher, P. Christ, D.~Hauff, J.~Ninkovic, F.~Petricca, 
F.~Pr\"obst, W.~Seidel, L.~Stodolsky}

\address{Max Planck Institut f\"ur Physik, M\"unchen, Germany}  

\author{F. v. Feilitzsch, T.~Jagemann, W.~Potzel, M.~Razeti, W.~Rau, M.~Stark, 
W.~Westphal, H.~Wulandari}

\address{Technische Universit\"at M\"unchen, Germany}  

\author{J. Jochum}
\address{Universit\"at T\"ubingen, Germany}

\author{C. Bucci}
\address{Laboratori Nazionali del Gran Sasso, Italy}

\maketitle
\abstracts{We present first competitive results on WIMP dark matter using the
phonon-light-detection technique.
A particularly strong limit for WIMPs with
coherent scattering results from selecting a region of
the phonon-light plane corresponding to tungsten recoils.
The observed count rate in the neutron band is  compatible with the rate
expected from neutron background. 
CRESST is presently being upgraded with a 66 channel SQUID readout system, 
a neutron shield and a muon veto system. This results in a significant 
improvement in sensitivity.
}

\section{Introduction}

Despite persuasive indirect evidence for the existence of dark
matter in the universe and in galaxies, the direct detection of
dark matter remains one of the outstanding experimental challenges
of present-day physics and cosmology. 
The Weakly Interacting Massive Particle (WIMP) is a well motivated
candidate for cold dark matter in the form of the lightest
supersymmetric particle. It is possible that
it can be detected by laboratory experiments, particularly using
cryogenic methods, which are well adapted to the small energy
deposit expected\cite{Goodman}.

\begin{figure}[t]
\centerline{
	\hspace*{-1cm}
	\epsfysize=5.5cm 
	\epsfbox{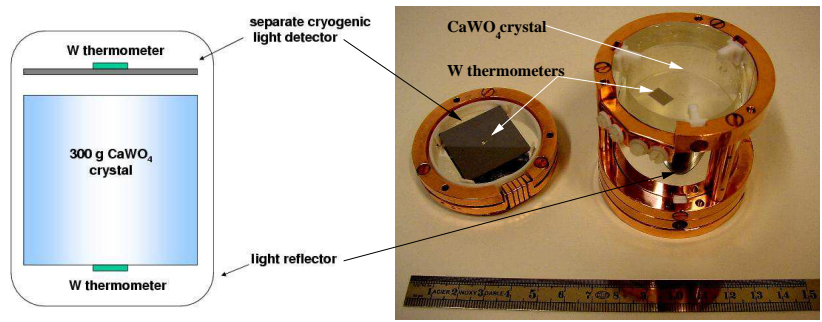}
	}
\caption{Left: Schematic sketch of the detector for
coincident phonon and light
measurement.
Right: Picture of a detector module, 
``phonon'' (right) and ``light channels'' (left).
\label{fig:detector_scheme}}
\end{figure}

\section{The CRESST experiment}\label{subsec:cresst}
In the CRESST experiment we attempt to
detect WIMP-nucleus   scattering  using cryogenic methods.
Results from the first phase of  CRESST using sapphire detectors have
been reported previously\cite{CRESST-I}.
For CRESST II\cite{proposal} we have  developed cryogenic detectors based on 
scintillating CaWO$_4$ crystals. When further equipped
with a light detector these
provide very efficient discrimination of
nuclear recoils from radioactive $\gamma$ and $\beta $ backgrounds,
down to recoil energies of about 10\,$~keV$. 
The mass of each crystal is about 300\,$g$.  
Passive background suppression is achieved with a low background
installation and the  deep underground  location at the Gran Sasso laboratory. 
The overburden of
3500 meter water equivalent reduces the
surface muon flux to about $1/(m^2 h)$. The  
detectors themselves are shielded against ambient radioactivity 
by low background copper and lead. 
A neutron shield and a muon veto, to be installed 
for CRESST-II, were not present for the data presented here.
A  four channel SQUID 
system allowed the simultaneous operation of only two phonon/light
modules. 

\subsection{Detectors}
A single detector module consists of a scintillating CaWO$_4$
crystal, operated as a cryogenic calorimeter (the ``phonon channel''),
and a nearby but separate cryogenic detector optimized for the
detection of
scintillation photons (the ``light channel'').  The phonon channel
is  designed to  measure the energy transfer to a nucleus
of the CaWO$_4$ crystal in a WIMP-nucleus elastic scattering. 
Since a recoil-nucleus differs substantially in the yield of scintillation
light from an electron or a $\gamma$-quanta of the  same energy, 
an effective background discrimination against $\gamma$-particles and
electrons is obtained by a
simultaneous measurement of the  phonon and light signals. 
The prototype detector modules  used here 
consist of a 300\,$g$
cylindrical CaWO$_4$ crystal with 40\,$mm$ diameter and height, and an
associated cryogenic light detector\cite{Meunier,LTD10}. 
The light detector is mounted
close to a flat surface of the
CaWO$_4$ crystal, and both detectors are enclosed by a 
housing made of a highly reflective polymeric multilayer foil. 
The arrangement is shown in
Fig.~\ref{fig:detector_scheme}.
The detectors are operated at a temperature of about $10\,mK$ where
the tungsten thermometer
is in its transition between the superconducting and
the normal conducting state, so that a small temperature rise of the
thermometer
leads to a relatively large rise of its resistance which is
measured by means of a two-armed parallel circuit. One branch 
of the circuit has the superconducting film and  
the other comprises a reference resistor in series with the input
coil of a SQUID,
which provides a sensitive measurement of current changes.
A rise in the thermometer resistance and so an increase in current through
the SQUID coil is then observed as a rise  in SQUID output voltage.  
Incoming pulses are recorded using a 16-bit transient recorder.
For a more detailed description of the experimental setup 
see\cite{astro_limit}.

\subsection{Temperature stability and energy calibration }\label{sec:stabil}
To monitor the long term stability of the thermometers particle-event 
like heater test pulses with a range of discrete energies 
in the energy
region of interest were sent every $30\,s$ throughout both dark matter
and calibration runs. The
response to these proved to be stable 
within the energy resolution of the detectors.
The accuracy of the energy calibration  from 10 to 40\,$~keV,$ as
relevant for the WIMP search is in the range of a few percent.
This can be inferred from a peak at 47.1\,$~keV$
which appeared in the energy spectrum of the phonon channel
with a rate of (3.2 $\pm$ 0.5) $counts/day$. We associate
this peak with an external $^{210}$Pb contamination 
resulting in a $\gamma$-line at
46.5\,$~keV$. The FWHM of the measured peak is 1.0\,$~keV$,   
identical with that for the heater pulses. This good resolution confirms 
the  stability of the response during the dark matter run.

\section{Results and discussion} \label{sec:results}

\begin{figure}[t]

\centerline{
		\epsfysize=6.2cm
		\epsfbox{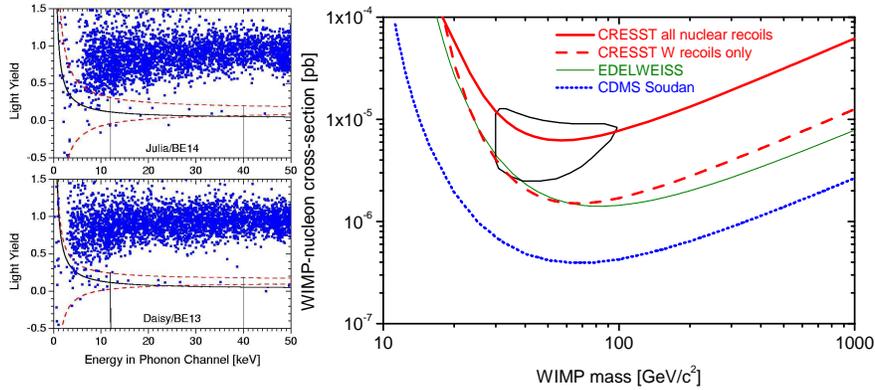}
	}   
\caption[]{Left: Low energy event distributions in the dark matter run 
for the two modules.
Right:  Coherent  scattering exclusion limits from the
dark matter run. The enclosed region represents the claim of a positive
signal by the DAMA collaboration\cite{DAMA}.}
\label{fig-ratio_12_34}
\end{figure}

The results shown here were obtained in measurements taken between 
January 31 and March 23, 2004. The total exposures 
after all cuts are $9.809~kg~d$ and $10.724~kg~d$ for the two modules.
The low-energy data from the dark matter run is presented in
Fig.~\ref{fig-ratio_12_34} as a scatter plot.
The determination of a nuclear recoil acceptance region in the phonon-light 
plane is based on a knowledge of
the ``quenching factor'', i.e. the reduction of the light output of a
nuclear recoil event relative to an electron/photon event of the
same energy. 
From earlier measurements of recoil neutrons a quenching factor of $Q$=7.4 was
determined\cite{Meunier}. From this and the light detector resolution the 
90\,\% acceptance band for nuclear recoils is calculated, which
is shown in the left panel of Fig. \ref{fig-ratio_12_34} 
as the area below the upper dashed lines.
If we attribute all 16 events from the two detector modules
in the acceptance region (corresponding to a count rate of 
$R=(0.87\pm 0.22) /(kg\,day)$ \;) to WIMP
interactions, we can set  a conservative upper limit for the WIMP 
scattering cross section shown as the full line in Fig. \ref{fig-ratio_12_34}.
For details on the assumptions made see Ref.\cite{astro_limit}.
Monte Carlo simulations for our setup without  neutron shield
yield an estimate for the
neutron background of about 0.6\,$~events/(kg~day)$ for $12~keV \leq
E_{recoil}\leq 40~keV$, in reasonable
agreement with the observed rate\cite{Hesti}.

Kinematic considerations
and simulations show that the contribution 
to the neutron spectrum is dominated by recoil of 
oxygen nuclei within the CaWO$_4$ 
crystal\cite{Hesti}. Hence the measured quenching factor 
will be mostly due to oxygen recoils.
However, if one assumes coherent scattering of WIMPs by the nucleus, the 
interaction cross section will be proportional to $A^2$. Thus WIMPs are mainly
expected to interact with tungsten nuclei.
In a seperate measurement the quenching factor of tungsten has been determined
to be between $Q_W \ge$40 \cite{jelena} for 18\,$~keV$ and 
36\,$~keV$ tungsten energies,
thus WIMP interactions are expected to lie in a seperate band below the
one determined from neutron recoils. The discrimination efficiency of the
two bands will strongly depend on the energy resolution of 
the attributed light detector. In Fig. \ref{fig-ratio_12_34} 
the 90\,\%
acceptance regions for recoil by tungsten are indicated by the 
area below the solid lines.
Since we know that the light detector of one of the 
two modules did have a slightly deteriorated resolution
we discard these data for the extraction 
of the WIMP coherent cross-section limit.
For the better of the two modules there are no recoil events  below 
the full line in the energy range from 12 to 40\,$~keV$.
Using these data to set a limit  we 
obtain   the thick dashed line in
the right panel of Fig.~\ref{fig-ratio_12_34}. 
As a check we lowered the threshold  
to 10\,$~keV$ to include the two events   
at 10.5\,$~keV$ and 11.3\,$~keV$ below the tungsten line
and obtained essentially the same curve.
At a WIMP mass of 60\,$~GeV/c^2$, these tungsten limits, which were
obtained without any neutron 
shielding, are identical to the limits set by EDELWEISS
\cite{Edelweiss}. 
Very recent results from CDMS at the Soudan Underground Laboratory
\cite{CDMS} have
improved these limits by a further factor of four.

\begin{figure}[t]
\centerline{
	\epsfysize=6.5cm\epsfbox{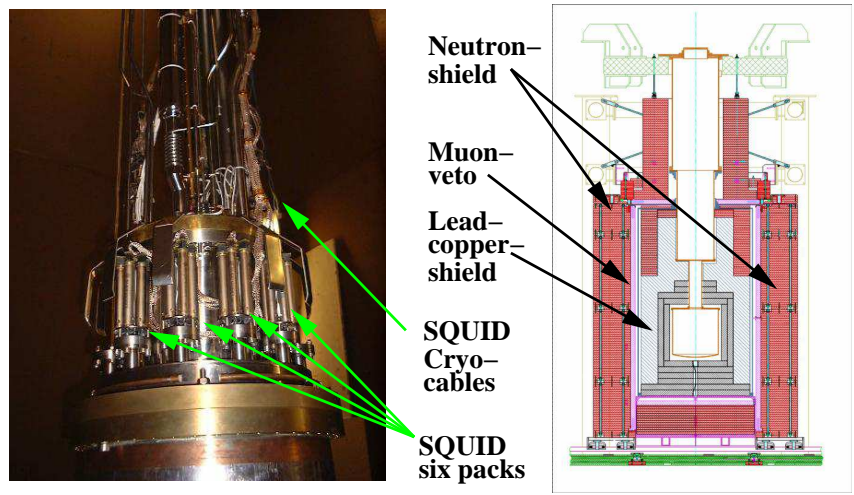}
	}   
\caption{Left: First 24 SQUIDs installed on the cryostat insert at 
Gran Sasso. Right: schematic view of CRESST cryostat with
neutron and muon shield.\label{fig:squids}}
\end{figure}

\section{Outlook: Upgrade with neutron shield and a 
66-channel SQUID readout system}
The limits presented in section~\ref{sec:results} were 
obtained using measurements
that were taken without a neutron shield.
Thus the sensitivity of these data is limited by the 
neutron background as well as by the limited exposure.
Presently the CRESST setup is being upgraded with a 66-channel
SQUID readout system\cite{squids}. This will allow the installation 
of 33 detector modules providing 10\,$~kg$
of target material. Together with the neutron shield that has been
installed in October 2004 and the muon veto
that is presently under construction this shoud improve CRESST's
sensitivity to WIMP dark matter interactions 
by two orders of magnitude. Fig.~\ref{fig:squids}
shows the first 24 SQUIDs that were installed onto the cryostat insert
and cooled down to liquid helium temperature
at the Gran Sasso underground laboratory. All the tested SQUIDs performed
well at liquid helium temperature.
The upgrade of the experiment will be completed in early 2005.


\begin{thebibliography}{0}


\bibitem{Goodman}
M.W.~Goodman and E.~Witten, {\it Phys. Rev.} {\bf D31}, 3059 (1985).
\bibitem{CRESST-I}
G.~Angloher et al.,
{\it Astropart. Phys.} {\bf 18}, 43 (2002).
\bibitem{proposal}
Proposal for a second phase of CRESST, {\it MPI-PhE/2001-02}.
\bibitem{Meunier}
P.~Meunier et al. {\it Appl.\ Phys.\ Lett.\ }  {\bf 75}, 1335 (1999).
\bibitem{LTD10}
G.~Angloher et al., {\it Nucl. Instr. Meth.} {\bf A 520}
(2004)
108-111. and F.~Petricca, et al., {\it Nucl. Instr. Meth.} {\bf A 520}, 193
(2004). 
\bibitem{astro_limit} G. Angloher et al., submitted to {\it Astropart. Phys.}, 
{\bf astro-ph/0408006}.
\bibitem{DAMA} R. Bernabei et al., {\it Phys. Lett.} {\bf B480}, 23 (2000).
\bibitem{Hesti} H.~Wulandari et al., in press at {\it Astropart. Phys.}, 
	{\bf hep-ex/0401032}. 
\bibitem{jelena} J. Ninkovic,  PHD thesis, to be published.
\bibitem{Edelweiss} A. Benoit et. al., Phys. Lett. {\bf B545} (2002) 43.
\bibitem{CDMS} D. S. Akerib et al.,  submitted to 
{\it Phys. Rev. Lett.}, {\bf astro-ph/0405033}.
\bibitem{squids} S. Henry et al. {\it Nucl. Instr. Meth.} 
{\bf A 520}, 588 (2004).

\end{thebibliography}
\end{document}